\documentstyle[aps,epsfig]{revtex}

\begin{document}

\title{  Peculiarities of structural changes in Pd and Pd-alloys.}
\author{V.M.Avdjukhina, A.A.Katsnelson~\thanks{E-mail:
albert@solst.phys.msu.su}, G.P.Revkevich, E.A. Goron}
\address{Department of Solid State Physics of Moscow State University,
Moscow ,Russia}

\maketitle

\begin{abstract} It has been discovered that kinetics of structural changes in Pd and 
Pd-alloys under the influence of dissolved hydrogen seems to be non-trivial.
As a rule, structural changes in these systems include 
correlation changes of 
co-existing phases concentration and the defect structure character.
  These changes have the stage character.  Basic stages are: incubation
 period, quick degassing, temporal stabilization and post-stabilization
 period.    The structural changes and phase transformation in Pd-M-H
 kinetics happens to be non-monotonous (oscillating or stochastic ).
 Such time dependence can be observed during tens thousand hours.
  The hypothesis based on the ideas of non-equilibrium dynamics and 
specific attributes of hydrogen interaction with matrix defect and 
atoms in Pd-systems has been offered in order to explain discovered phenomena.
\end{abstract}

\section{Introduction}
Unique ability of palladium for big quantity of hydrogen dissolving was
discovered about 130 years ago by Graham [1], the one of  most known UK
chemists. The dissolving of H in Pd considerably has influence over physical
properties [2,5]. The Pd-H alloy is diamagnetic and super conductive though
Pd is strongly paramagnetic. These differences are connected with the  atomic
 and electronic structural changes at H dissolving in Pd. Here only atomic
structure characteristics will be considered.
Solid solutions of H in Pd create $\alpha$ - phase if H concentration ( this
values are described as $n_H/n_{Pd}$  where $n_H$  and $n_{Pd}$  are quantities of H
 and Pd atoms,respectively ) is not more than 0.02-0.03 or $\beta$- phase if H
 concentration is more than 0.6.  If H concentration is between those values
mix of these
phases appears. In both phases Pd - atoms create fcc-structure, where H -
atoms are in  octapores. Distance between Pd - atoms in $\beta$ - phase exceeds
corresponding distance in $\alpha$ - phase on 3\%`. That is why $\alpha\leftrightarrow\beta$
phase conversations processes are followed  by  the defect generation.
Structural changes taking place in Pd - H during the saturation and following
degassing have become an object of systematic researching  very recently. The
results of the researching of $\alpha$ - and $\beta$ - phases peculiarities
 changes of  shares relations  and defect structures forming during
 $\alpha\rightarrow\beta$ ( during saturation) and during $\beta\rightarrow\alpha$
(during degassing) processes were the most interesting. These processes
 including their kinetics have been fully discovered in the cycle of our
researching [4 - 12].

Non-ordinary data received in these researching showed the important role of
 defect structure forming    ( at saturation ) and transforming ( during degassing ).
That is why the character of hydrogen influence on structural changes in Pd - H solutions
 and their kinetics were very interesting.  Disordered mutual Pd and M atoms
 distribution is the source of the additional defect structure formation
comparing to  Pd. The brightest effects should be expected in those solutions
 that are different from Pd by the degree of  affinity to hydrogen.  It could
 lead to  some new characteristics of structural changes in palladium systems
appearance under the influence of the saturated hydrogen.    The results of
the researching [13-16] confirmed that the structural changes in palladium
solutions have non-trivial character.

The purpose of this paper is to consider Pd and some Pd-M solutions structural
 changes characteristics at the saturation and following degassing.  The
researching was led by X-rays methods that are described in detail in [4-16].

The hydrogen saturation has been led by electro-chemical method at the
current density 2.5-80mA/sq.~sm and the saturation time was 7-120min.
 The samples degassing going spontaneously while kept at the open air
 leads to spontaneous structural changes in the samples saturated
by hydrogen.

Researched samples were the plates with the thickness of 0.1-2mm,
 whose surfaces were polished and ground.  The deformation effects in some
samples were taken off by the annealing.  The saturation was led either by
one act or by some acts of  repeating cycles 'saturation-degassing' (cycling).

\section{  The peculiarities of phase transformation in Pd-H.}

Let's take a look at the curves of the $\beta$-phase concentration p(t) dependence
on the saturation time at  low current density (2.5mA/sq.sm). Fig. 1 shows
the dependence of lnp(t) function on the saturation time. Curves 1,2,3 relate
to such blocks of coherent scattering (further - hkl-blocks), whose
crystallographic plains with (100), (311), (110) indexes, respectively,
are parallel to external surface of the sample.  Fig.1 shows also that the
experimental points lay on the straight lines that do not go through the
coordinates beginning.

Received data shows the dependence of $\beta$ - phase concentration on time,
which is

\begin{equation}
p(t)=1-\exp[-\gamma(t-t_0)]
\end{equation}

 here  $\gamma$- is
logarithmic velocity of $\beta$ -phase quantity growth, $t_0$-the incubation
 period duration.  The logarithmic velocity of the growth and the incubation
period duration   depend on crystallographic orientation of  block plains
relatively to external surface. $\gamma$ reaches its maximum at ( 100 ) blocks
 and minimum at ( 110 ) blocks,$t_0$ , on the contrary, reaches its minimum
at (100) blocks and its maximum at (110) blocks. The orientation dependence
considered to be stronger for $\gamma$ value than for $t_0$ 's. Besides $t_0$
 and $\gamma$ seemed to depend on current density $j$  essentially. At  $j$
increasing up to 25mA/sq.cm $t_0$ becomes almost 40 times less for ( 100 )
blocks,$\gamma$  becomes one order more.

This data could be explained basing on the  phase transformation of 1st order
 [ 7 ] kinetics theory. According to this theory phase transformation from
 $\beta-$ to $\alpha$ - phase takes place then the decreasing of volume energy
 connected to creating of new phase exceeds energy loss which has gone for
 appearance of boundaries  between new phase and old one, defect generation,
increasing of  matrix elastic energy connected to elastic tension appearance
because of specific volume phase difference. During $\alpha\rightarrow\beta$
- transformation effective pressure of saturating hydrogen, defined by current
 density in electrolytic bath  is included in thermodynamic stimulus. The
embryos of new phase at $\alpha\leftrightarrow\beta$  transformation have
 the plate form. That is why energetic loss connected to elastic tension  in
 matrix because of the inbuilding there of the new embryos  has an anisotropy .
 Elastic energy of this kind of embryos appearing reaches minimum if their
surfaces are parallel to the crystallographic plains kind of ( 100 ).

Defect structure has an influence over thermodynamic stimulus of phase
 transformations and kinetics of this transformation.  It has been shown
experimentally that the phase transformation takes place by the way
of spontaneous moving of the boundary between $\alpha$  and $\beta$ - phases.
Because of that the velocity of $\beta$  - phase concentration increasing
 depends on energetic barriers value, which have to be got over during
this boundary migration on the potential relief of $\alpha$ -phase crystal.
  The defects leading to non-regularly distributed energetic barriers of
 different highness hamper the board moving and decreasing of $\alpha\rightarrow\beta$
  transformation velocity .  The energy of  inter-phase boundary migration
 reaches its minimum if the embryo surface is parallel to the crystallographic
plains kind of (100).

Specified factors explain the existence of the incubation period and its
anisotropy,  decreasing $t_0$ at the j increasing and its increasing at the
defect concentration increasing, stronger anisotropy of new phase growth
velocity.

Let us consider the peculiarity of the structural changes during $\beta\leftrightarrow\alpha$
   transformation.
  The researching was led after one-act saturation of  annealed and deformed
samples and annealed samples saturated by a few cycles.

The $\beta\rightarrow\alpha$ transformation in one act saturated ( j=40mA/sq.sm,

$t_{sat}$ =15min) annealed sample had started right away after the saturation,
 during first 25 hours the $\beta$- phase contains became 30 times less.
 First 5 hours had been an incubation period in deformed sample.
 The $\beta$-phase concentration had became 2.5 times less during next 25 hours
. During next 150
hours it had decreased to  30 \% from the original value.

The $\beta$ - phase concentration changing at cycle saturation of  annealed
sample is showed by Fig. 2 and 3 [6,11]. Fig. 2 shows that the incubation
period is absent during first three cycles. The degassing velocity is
decreasing while the number of cycles is
increasing. The incubation period appears after fourth saturation. The curves
of dependence of  the $\beta$-phase concentration after ninth saturation are
showed on Fig. 3. Right after this saturation p (t) was 80\%. This value
stays unchangeable during 4 thousand hours. After that the $\beta$ -phase
concentration was decreasing during 46 thousand hours. Then the concentration of this phase
stopped changing. This stop was fixed during 50 thousand hours. After that
the  p (t) changing which has oscillating character starts again.

Therefore, $\beta\rightarrow\alpha$  process has the stage character. Unlike
the $\alpha\rightarrow\beta$  process, the $\beta\rightarrow\alpha$
transformation takes place
spontaneously  while kept at open air.  Its initial  stage is  hampered
by original defect structure. Further process developing is specified by  new
defect
generation and following defect structure transformation.

Regarding to it the incubation period was noticed after  one-act saturation
only in deformed sample. Starting after fourth saturation it was noticed in
annealed sample. After ninth saturation its duration was much longer than
after the mechanical treatment applied to surface because the density of
dislocations is essentially bigger .

The p(t) decreasing during next stage of degassing corresponds to the
exponential characteristics only at the defect concentration which is not big.
  The increasing of this concentration and the creating of complexes (vacation
 complexes, dislocation walls ) following it, leads to appearance of  wider
and higher energetic barriers in a space.  It provides, regarding to [8],
its transition to power and even to logarithmic dependence of p(t) decreasing.
  The transition to the next stage where p(t) stops changing is caused by
specific transformation of defect structure.  One of its characteristics is
blocks growth in $\alpha$-phase [ ], which takes place because of dislocation
 walls migration to blocks boundaries. Because this process happens in
different parts of sample with different velocities, the additional
non-regularity appears in the system in distribution of energetic barriers on
potential relief of the $\alpha$-phase crystals.    This leads to additional
hampering of inter-phase boundaries, which stops the process of p(t)
decreasing. After the blocks enlargement process finishing this hampering
factor of boundaries migration disappears.  The defect and hydrogen
accumulation in blocks boundaries can lead to opposite procces.  As a result
the p(t) changing procces, which is going to have an oscillating charactert
may go on  This leads to the next stage of  relaxation.

  Therefore, the character of the structural changes at hydrogen saturation
and degassing seemed to be closely connected to the defect structure
transformation process taking place simultaneously.

\section {The structural changes in Pd-M-H alloys and their time dependence.}
Found non-ordinary kinetics of structural changes in Pd-H caused by
defect structure transformation had stimulated  the structural changes
kinetics researching in Pd-M-H alloys.

The kinetics aspects of those changes were being researched on Pd alloys with
W[13-14], Sm[15], Er[16-21] and,besides,the most in detail on annealed
Pd-W (11.3 at\% W) and deformed Pd-Er (8at\%Er).  The ingredients of those
alloys are essentially distinguished by the degree of the affinity to hydrogen
and,besides,W's affinity to hydrogen is lower than Pd's and Er has higher one
than Pd does.  Regarding to the diagrams of equilibrium for these alloys, $\beta$
-phase does not appear.  Pd-W alloy (11.3at\%W) in accordance to [23] is
characterized by the appearance of regions rich in W, that sizes are 2-3nm
and superfluous concentration of  vacancions [24].  Pd-Er alloys (8at\%Er) is
close to the solubility boundary.

Let us take a look at the experimental data for Pd-W alloys. Fig. 4 shows the
dependence of $ln(I_{400}/I_{200})$  on time after third saturation (here $I_{400}$
  and $I_{200}$ are normalized intensities of the diffraction maximums (400) and
 (200)[13-14]. This dependence  seems to be quasi-periodic.Its character shows that
the oscillation processes of two types take place in the system. One of them
is connected to the structural changes causing the quasi-periodic changes
of $ln(I_{400}/I_{200})$  function.  On the initial stage those oscillation
have a period of 7 days approximately.  The second one is connected to
the structure changes that are characterized by abrupt short-term  enbroadings
 of diffraction maximums and decreasing of their top parts, that happen every
4-5 weeks ( the enbroaded diffraction maximums 'wings' expand so far that
their proper measurements are unlikely possible).  Quasi periodic structural
changes of 1-st type transform to the stochastic ones after system's
transition through the structural change of 2-nd type.

The $ln(I_{400}/I_{200})$ function decreasing  for the first of those 
processes can be caused by the appearance of defect regions that sizes 
are 2-3 nm[25].
  Those regions differ from the matrix by the specific volume.  The increasing
 of  function can be connected either to dissolving of those regions or  to
the approaching of their specific volume to the matrix's one ( their
disappearance as defect ones). The width oscillations of diffraction maxima
can be connected to appearance and following disintegration of  dislocation
loops that sizes are 5-10nm[25].

The oscillation structural changes appearance
after the hydrogen saturation of the system indicates the clusters rich in
hydrogen formation in the matrix rich in Pd.  The specific volume of these
clusters is bigger than matrix's and because of it  they are not stable in
thermodynamics meaning in normal conditions.  For the non-contradictory model
 of this phenomenon construction one should suppose that the non-stability
decreasing would take place in account of its defect degree decreasing because
 of the diffusion of the superfluous vacancies being in this alloy into
clusters enriched by hydrogen.  Vacancies deficit arising during this process
in the regions rich in W will be stimulated by the contra-directed vacancies
diffusion.  The consequence of it will be the oscillation process of vacancies
moving that will lead to the structural oscillation of 1-st type.  After a few
 periods the concentration of vacancies in regions rich in palladium become
so high that it seems more profitable to form large vacancies dislocation's
loops during the intermediate period.  But those loops are not stable.
Soon after their arising they disintegrate, after that the process of defect
clusters formation and disappearance renews.  The offered model is, of course,
 hypothesis and needs more direct proves.  Nevertheless, it describes
experimental data and, undoubtedly, can be the base of stricter model for the
found phenomenon researching.

The initial condition of deformed by grinding and polishing Pd-8\%atEr alloy
is characterized by essentially non-homogeneous distribution of the
ingredients and availability of strong enough stretching tension which is
perpendicular to the surface. After the hydrogen saturation this tension
transformed to compression tension as strong as it was. Maximum compression
tension reaches in 2 days after the hydrogen saturation.  After 8 days its
value has become 25 \% less, after that it stays practically the same during
1.~5 years of the observation[17-18].

The form of the diffraction maxima had changed after the saturation.
They became doublets (fig.5).  This indicates the arising of  two phases
that essentially differ from each other by the period of lattice in
considered system.  One may see that the profiles of maxima dependence on
time has oscillating  character (fig.6).  The computer analysis of those
profiles lets us determine the time dependencies of differences in
concentrations of erbium in corresponding phases and specific portions of
those phases.  The data presented on the fig.7 shows that non-regular
oscillating (stochastic) changes of the indicated characteristics have been
taking place for 1.5 years.  These oscillations happen on the initial stage of
 degassing when there is 10-20\% of  H in the system and on the later stages
when the concentration of H is not more than 1\%.

For the received data explaining the model taking into consideration the idea
of microscopic theory of alloys and synergetic has been elaborated.
The lattice compression which has been found after the saturation can be
caused by the transformation of defect-metal (DM) complexes being in alloy
before the saturation into hydrogen-defect-metal (HDM) complexes because of
the bigger energy of  hydrogen -defect bond in Pd[26].  Therefore, the last
complexes have low specific volume.  These complexes because
of  high erbium's affinity to hydrogen attract erbium atoms and become traps
for erbium.  These traps play mutual role.  They keep the system in
non-equilibrium condition and lead to rising diffusion appearance.  According
to synergetic conception, non-equilibrium condition of the system admits the
appearance of the oscillating processes [27] connected to self-organization of
 defect-structural states.  The competition of rising and gradient diffusions
present the mechanism allowing possible oscillating processes to be realized.
  The diffusion fluxes  multitude caused by this competition in two-phase
system containing the Er traps leads to oscillating character of the phase
transformation kinetics.  The stochastic character of those oscillations is
connected to the essential difference of the relaxation time of different
oscillating processes in considered system.

\section{Conclusion.  }

Therefore, the structural changes peculiarities in hydrogen-containing systems
 Pd-H and Pd-M-H, whose M-ingredients essentially differ from each other by
the degree of affinity to hydrogen were considered in this paper.  Non-trivial
 kinetics of $\beta\rightarrow\alpha$ transformation has been found for Pd-H.
  The most important its peculiarity is the alternation of the stages where
the  $\beta$-phase concentration changing happens or does not happen  This
phenomenon is connected to the influence of as original defect structure as
its  transformation during the procces of  phase transformation
over $\beta\rightarrow\alpha$ transformation kinetics.  The second important
$\alpha\leftrightarrow\beta$  transformation peculiarity is strong
dependence of the p(t) function on (hkl)-blocks orientation relatively its
 external surface.  It caused by the elastic energy dependence evoked by the
plate formation, on its orientation in the blocks.

The structural changes kinetics in Pd-M-H has an oscillating character.  This
kinetics peculiarities are connected to the defect subsystem caused by
non- homogeneous distribution of the metal ingredients and ,as a result,
by hydrogen non-homogeneous distribution.   Hydrogen is being captured by
regions  that have high energy of the bond to hydrogen that keeps the system
 in non-equilibrium state. This  leads to diffusion fluxes multitude where
arising and gradient diffusion are competing.  As a result, 'static'
non-stability transforms into dynamics one similar to Benard's cells [27].
 According to real alloy structure the oscillation character of the structural
 changes can acquire different characters. The structural changes are the
defect regions appearance and disappearance alternation in Pd-W-H.
The structural changes have the form of phase transformations of stochastic
type in Pd-Er-H.

The long-term oscillating structural changes in hydrogen-containing alloys
correlate with the similar strength characteristics changes; for example
roll steel [28].  That is why actual direction of further investigations
is founding of such metallic alloys for that this phenomenon is observed,
researching of their nature and their practical use possibility.

Acknoledgements

This work has been supported by the Russian Fund of Fundamental Research,
under grant \#99-02-16135.

\begin{figure}  \label{fig1}
\epsfig{file=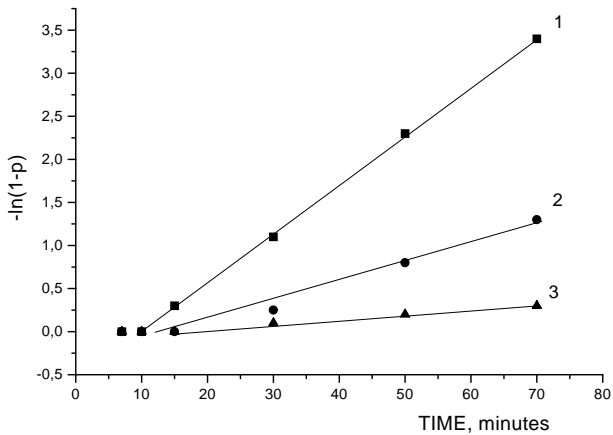,width=10cm}
\caption { Logarithmic dependence of $\beta$-phase concentration p on
 time after the hydrogen saturation for hkl-blocks:1-(100), 2-(311), 3-(110)}
\end{figure}

\begin{figure}  \label{fig2}
\epsfig{file=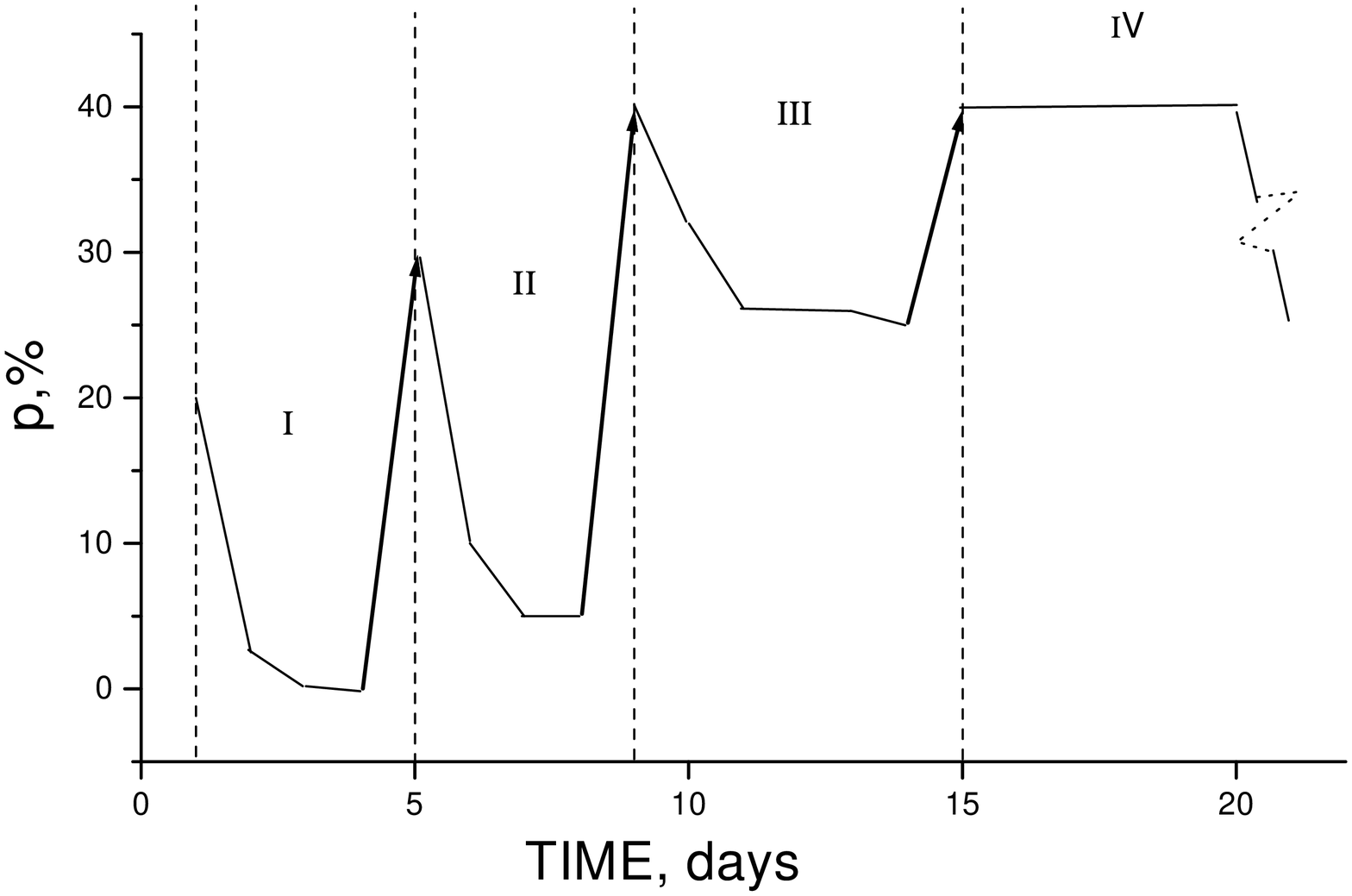,width=10cm}
\caption { $\beta$-phase concentration dependence on time for cycles from
 first to fourth;1,2,3 and 4 are cycles  numbers, the arrows show the p 
changing at hydrogen saturation}
\end{figure}

\begin{figure}  \label{fig3}
\epsfig{file=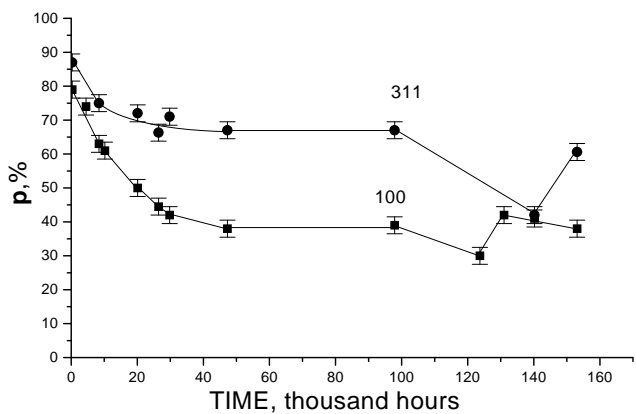,width=10cm}
\caption { $\beta$-phase concentration dependence on time after ninth 
hydrogen saturation for (100) and (311) blocks; incubation period is omitted}
\end{figure}

\begin{figure}  \label{fig4}
\epsfig{file=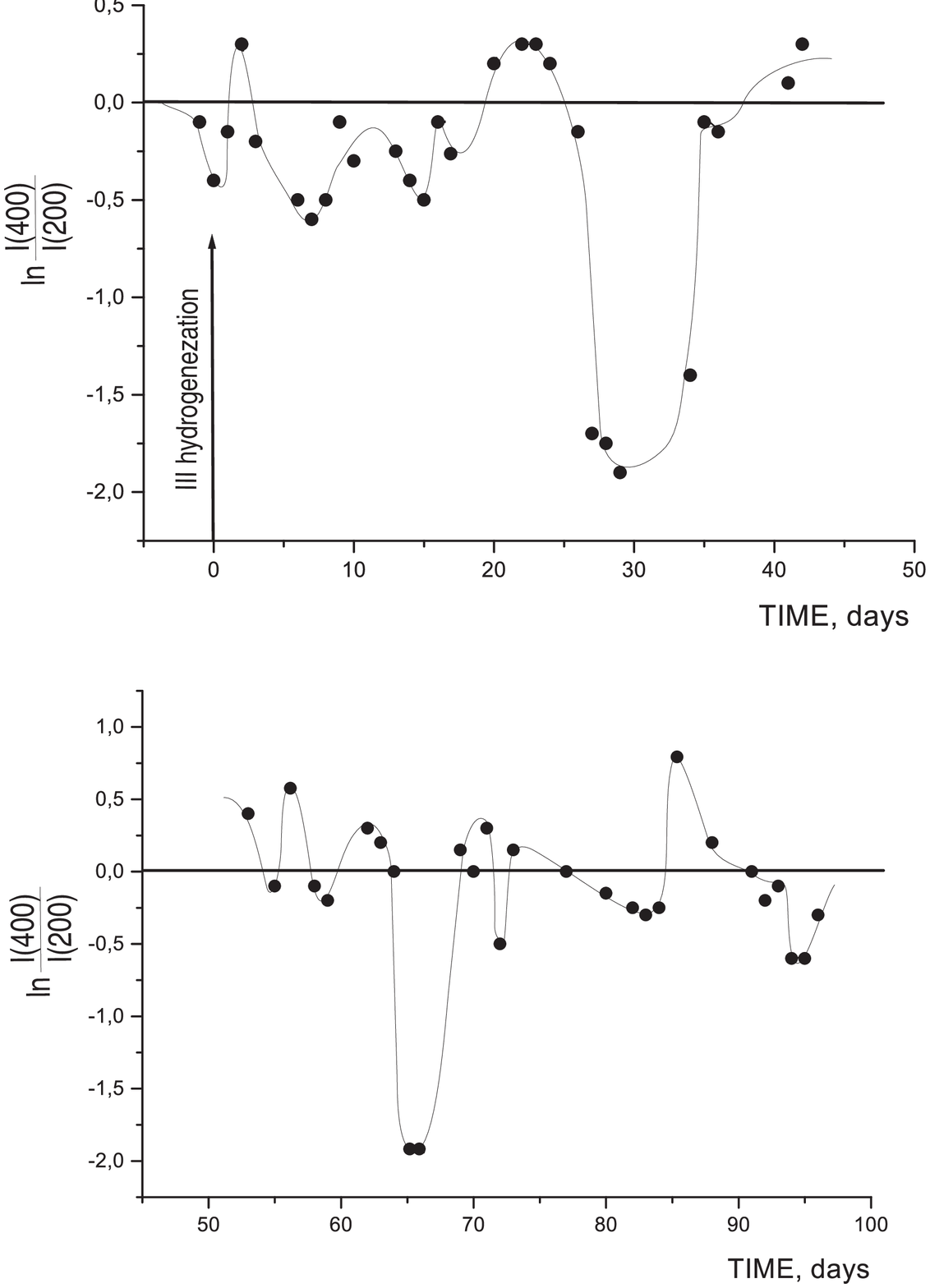,width=10cm}
\caption { ln($I_{400}$/$I_{200}$) dependence on relaxation time for 
Pd-11.3at\% W alloy after the third hydrogen saturation}
\end{figure}

\begin{figure}  \label{fig5}
\epsfig{file=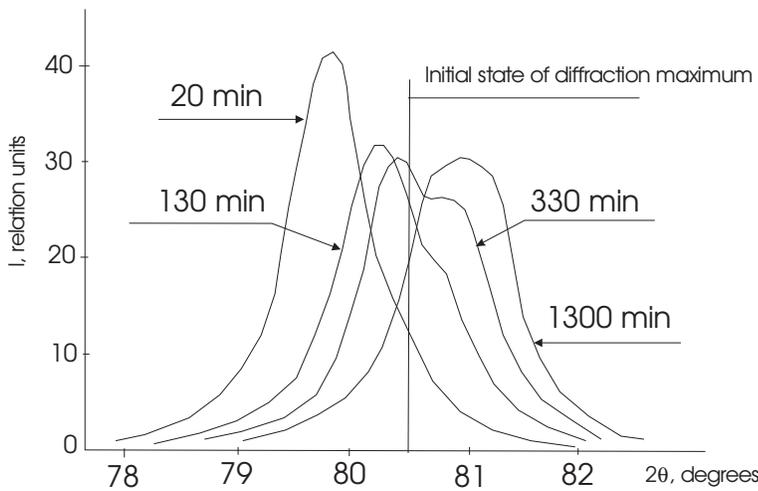,width=10cm}
\caption { The profiles of diffraction maximum for Pd-8at\%Er 
alloy 20 min, 130 min,330 min and 1300 min after the hydrogen saturation}
\end{figure}

\begin{figure}  \label{fig6}
\epsfig{file=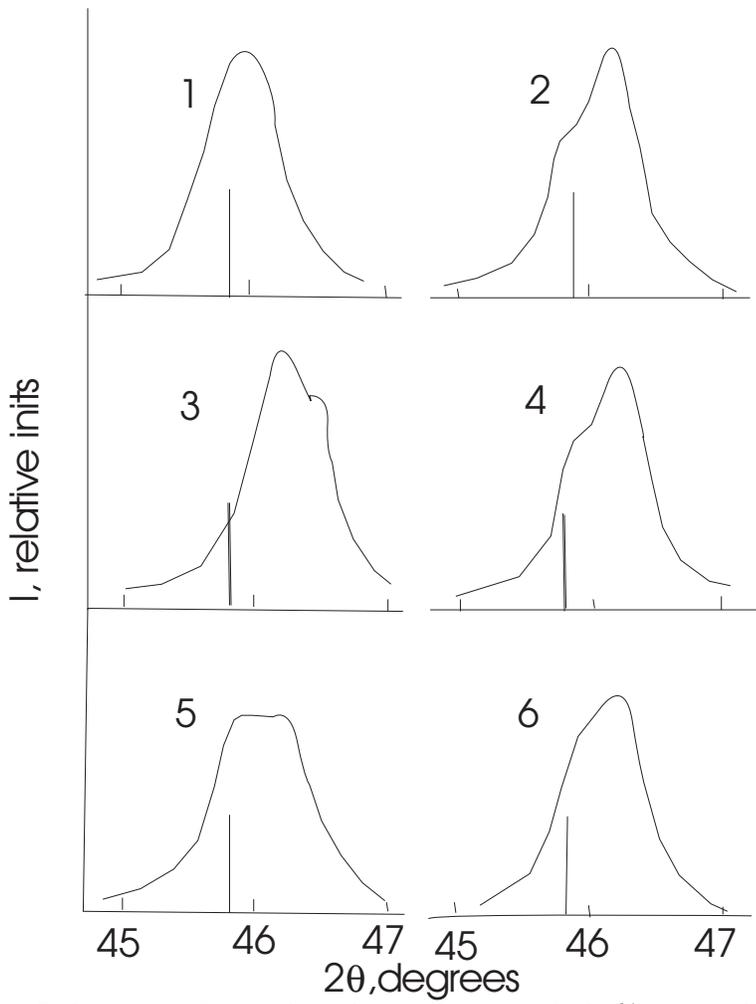,width=10cm}
\caption { The profiles of diffraction maximum for Pd-8at\% Er alloy 1.5 
  hours (1), 7 hours(2), 25 hours(3), 48 hours(4), 120 hours (5), 
4200 hours (6) after the hydrogen saturation}
\end{figure}

\begin{figure}  \label{fig7}
\epsfig{file=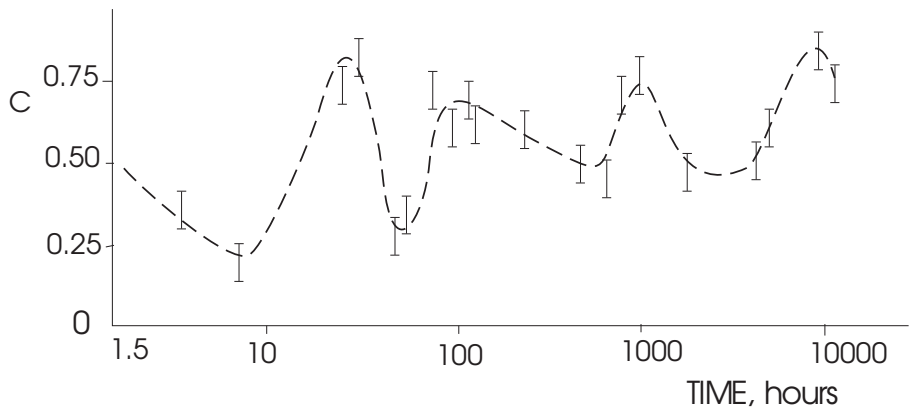,width=12cm}
\caption { Phase rich in Er concentration dependence on time for 
Pd-8at\%Er afte the hydrogen saturation}
\end{figure}

\end{document}